\documentstyle[12pt]{article}
\begin{document}
\begin{center}
{\bf{\Large Non-polarizing two--atom interaction potential
in the liquid $^4$He}}

\vspace{0.7cm}

{\large\bf D. B. Baranov $^{a}$, A. I. Kirillov $^{b}$,
V. S. Yarunin $^{a,}$
\footnote{Corresponding author. E-mail: yarunin@thsun1.jinr.ru}}

\vspace{0.7cm}

{\it $^a$ Joint Institute for Nuclear Research, Dubna 141980, Moscow Region,
 Russia}\\

{\it $^b$ Moscow Power Engineering Institute, Krasnokazarmennaja 14, E-250,
Moscow 111250, Russia}

\vspace{0.7cm}

\begin{quote}
A formula for potential $U$ of the $^4$He--$^4$He interaction in the
liquid state is obtained
by the direct electromagnetic computation as a function of the
interatomic distance  $R$. The potential decreases exponentially
at  large $R$. The further development and application of the result
are discussed.

\vspace{0.5cm}
\end{quote}
\end{center}

\vspace{0.2cm}

It is known, that the Leonard--Jones (L--J) potential
describes the interaction of  atoms of simple non-polar gases \cite{one}.
However, at  large
distances between atoms the interaction forces  fall faster, than it is
described by the L--J potential \cite{one}.
The corresponding corrections (including light atoms) are
suggested in papers \cite{three}--\cite{seven}.
These authors had a suspicion, that,
for light atoms, either L--J formula or pair interaction approximation,
or both of them are not valid. In this paper, we show that
the rapid decrease of the
potential can be explained, if we assume that the atom polarization is
not important in the liquid.

In this paper, we calculate the potential $U$ between charge densities of
electrons and nuclei  for $^4$He in the ground state. The interaction
energy of  two charge distributions with the densities
$\rho_1 (\bf{r})$ and  $\rho_2 (\bf{r})$ is
\begin{equation}
\label{one}
U=\int\frac{\rho_1 (\bf{r})\rho_2 (\bf{r' }) d\bf{r}d\bf{r'}}
{| \bf{r}-\bf{r'}|},
\end{equation}
where the charge densities are
\begin{equation}
\begin{array}{l}
\label{two}
\rho_1 (\bf{r})=q\delta (\bf{r}-\bf{r_1})-q P(\bf{r}-
\bf{r_1}), \\
\phantom{ffff}\\
\rho_2 (\bf{r'})=q\delta (\bf{r'}-\bf{r_2})-q P(\bf{r'}-
\bf{r_2}).
\end{array}
\end{equation}
Here $P(\bf{r}-\bf{r_1})$ and $P(\bf{r'}-\bf{r_2})$ are
the electron probability densities at the points $\bf{r}$ and $\bf{r'}$,
the atom nuclei being at the points
$\bf{r_1}$ and $\bf{r_2}$,  and  $q=2e$.
 Using (\ref{two}) in (\ref{one}),  we have
$$
\int\frac{\rho_1 (\bf{r})\rho_2 (\bf{r'})d\bf{r}d\bf{r'}}
{| \bf{r}-\bf{r'}|}
=q^2\int\frac{\delta (\bf{r}-\bf{r_1})
\delta (\bf{r'}-\bf{r_2})d\bf{r}d\bf{r'}}{| \bf{r}-\bf{r'}|}-
$$
$$
-q^2\int\frac{\delta (\bf{r}-\bf{r_1})
 P(\bf{r'}-\bf{r_2}) d\bf{r}d\bf{r'}}
{| \bf{r}-\bf{r'}|}-
q^2\int\frac{\delta (\bf{r'}-\bf{r_2}) P(\bf{r}-
\bf{r_1}) d\bf{r}d\bf{r'}}{| \bf{r}-\bf{r'}|}+
$$
\begin{equation}
\label{eq2}
+q^2\int\frac{P(\bf{r}-\bf{r_1})
P(\bf{r'}-\bf{r_2})d\bf{r}d\bf{r'}}
{| \bf{r}-\bf{r'}|}=U_{nn}+U_{nc}+U_{nc}+U_{cc}
\end{equation}
It is evident, that the first term  $U_{nn}$ in  (\ref{eq2}) gives
the interaction
energy of two positive point-like charges of nuclei after
the integration over coordinates $\bf{r}$ and  $\bf{r'}$:
$$
U_{nn}=\frac{q^2}{R}.
$$

The term $U_{nc}$  describes the interaction between nucleus of
an atom and electron cloud of another.  We have
\begin{equation}
\label{nc}
U_{nc}=-q^2\int\frac{\delta (\bf{r}-\bf{r_1})
 P (\bf{r'}-\bf{r_2}) d\bf{r}d\bf{r'}}
{| \bf{r}-\bf{r'}|}=-
q^2\int\frac{ P (\bf{r'}-\bf{r_2}) d\bf{r'}}
{| \bf{r_1}-\bf{r'}|}=-
q^2\int\frac{ P (\bf{r_0}) d\bf{r_0}}
{|\bf{R}-\bf{r_0}|},
\end{equation}
where $\bf{r_0}=\bf{r'}-\bf{r_2}$ and $\bf{R}=\bf{r_1}-
\bf{r_2}$.
We assume that the electron probability density $P(\bf{r_0})$ is
spherically symmetric, the center being  at the nucleus (the accuracy of
this approximation is discussed at the end of the paper).
Integration with respect to the  spherical coordinates $\theta$ and
$\varphi$ in (\ref{nc}) gives
\begin{equation}
\label{cn1}
U_{nc}(R)=-\frac{q^2}{R}+
4\pi \frac{q^2}{R}\int\limits_{R}^{\infty}P(r_0)(r^2_0-Rr_0)dr_0
\end{equation}
We used the change of variable
\begin{equation}
\label{cc1}
\theta\to y=\sqrt{R^2+r_0^2-2Rr_0\sin\theta}, \quad
\int\limits_{0}^{\pi}\sin\theta d\theta\to\int\limits_{|R-r_0|}^{R+r_0}
\frac{ydy}{Rr_0}.
\end{equation}

The cloud--cloud interaction term $U_{cc}$ in (\ref{eq2}) after
the similar integration with respect to spherical coordinates
becomes
\begin{equation}
\label{cc}
U_{cc}=-\frac{2\pi}{R}\int\limits_{0}^{R}U_{nc}(x)\int\limits_{R-x}^{R+x}
P(y)ydy xdx-\frac{2\pi}{R}\int\limits_{R}^{\infty}U_{nc}(x)
\int\limits_{x-R}^{x+R}P(y)ydy xdx.
\end{equation}
The formulas for $U_{cc}$ and  $U_{nc}$ are valid for any
atoms in liquids and gazes at sufficiently low temperatures.
Concrete expressions
for $P(r)$ have to be used in (\ref{cn1}) and (\ref{cc}) to
obtain the final formula.  In particular, for
$^4$He, both electrons are at the state with
$n=1,\quad l=0$ (para--state). To obtain a simple formula for using in
preliminary computations, we may ignore the electron--electron interactions
in the cloud. Then the wave function of the electrons in the
$^4$He ground state is
$$
\Phi (r_3, r_4) =\Psi (r_3)\Psi (r_4)\chi_p,
$$
where $r_3, r_4$ are the electron coordinates,
$$
\Psi (r) =\frac{1}{\sqrt{\pi a^3}}\, e^{-r/a},\quad
a=\frac{\hbar^2}{Z e^2 m},\quad
\chi_p=\frac{1}{\sqrt{2}}[ | \uparrow \downarrow \rangle -| \downarrow
\uparrow \rangle ],
$$
$\chi_p$ is antisymmetric spin wave function of the para--state.

The product $\Psi (r_3)\Psi (r_4)$ describes the symmetric radial part
 of the ground--state wave function.
$|\Phi (r_3, r_4)|^2$ is the probability density to find one electron
at the point $r_3$, and another at the point $r_4$.
Because $|\chi_p |^2=1$, the probability density is
$$
|\Phi (r_3, r_4)|^2 = \Psi^2 (r_3)\Psi^2 (r_4).
$$
The charge density created by the two-electron cloud at the point $r$ is
$$
q P(r) = e |\Psi (r) |^2 +e |\Psi (r) |^2 =
q |\Psi (r) |^2.
$$
Using this formula and (\ref{cn1}), (\ref{cc}), we obtain
the expressions for $U_{nc}$ and $U_{cc}$:
\begin{equation}
\label{unc}
U_{nc}=-\frac{q^2}{R}+ q^2\left(\frac{1}{R}+\frac{1}{a}\right)e^{-2R/a}.
\end{equation}
\begin{equation}
\label{ucc}
U_{cc}= \frac{q^2}{R}-q^2\left(\frac{1}{R}+
\frac{1}{a}\right) e^{-2R/a}
-\frac{3q^2}{8a} e^{-2R/a}-\frac{q^2 R}{a^2}
\left( \frac{R}{6a}+\frac{3}{4}\right) e^{-2R/a}.
\end{equation}
Thus, the interaction potential between two atoms at a distance $R$ is
\begin{equation}
\label{pot}
U(R)=U_{nn}+U_{nc}+U_{nc}+U_{cc}=
\frac{q^2}{a}\left[ -\frac{1}{6}{\left( \frac{R}{a}\right)}^2
-\frac{3}{4}\left(\frac{R}{a}\right)+\frac{5}{8}+\left(\frac{a}{R}\right)
\right] e^{-2R/a}.
\end{equation}
It is obvious, that
$$
U(R)\simeq -\frac{q^2 R^2}{a^3} e^{-2R/a}, \quad \mbox{ as } R\gg a.
$$
The exponential decrease of our potential $U$ contrasts with
$R^{-6}$--decrease in other formulas for $^4$He--$^4$He potential.
The slow $R^{-6}$--decrease is provided by the use only the dipole
interaction is taken into account in these formulas.

Our formula (\ref{pot}) for $U(R)$ does not contain arbitrary parameters.
It is approximate, however, because we neglected the electron-electron
interaction for obtaining $\Psi (r_3, r_4)$. Nevertheless, formula suggest
an idea, that the semi-phenomenological potential of the form
\begin{equation}
\label{pot1}
U(R)=\left[A{\left(\frac{R}{a}\right)}^{2}+B\left(\frac{R}{a}\right)+C
+D\left(\frac{a}{R}\right)\right] e^{-\gamma R/a}
\end{equation}
is of interest for the theory of the liquid
$^4$He maxon--roton spectrum, that considerably depends on the form of
interatomic potential \cite{eight}.

To take the electron interaction in the cloud into account,
we have to  use some realistic wave functions of the $^4$He electrons in the
ground state (see, e. g.,  \cite{nine}).
For instance, the wave function of Hylleraas
with three parameters gives additional terms up to $R^{6}$ inside the
square brackets in (\ref{pot1}).
We emphasize that our formulas are designed for describing the two--atom
interactions in the liquids and gases and we suppose, that no polarization
effect exists for any pair  of atoms in liquid due to its compensation.

The interaction potential of two atom in the vacuum  in general has a
contribution caused by the atom polarization.
The atom polarization results in that the centers  of electron clouds
do not coincide with the nuclei and the clouds are deformed.
The first estimate of a polarization effect can be easily described
by our formulas
with $R$ in $U_{nc}$ less than $R$ in $U_{nn}$ and $U_{cc}$. In such
case, taking $\Delta$ as a shift of a positive and negative charge centers,
we have a Buckingham--type formula
\begin{equation}
U(R)=U_{nn}(R)+2U_{nc}(R-\Delta )+U_{cc}(R)=-\frac{2q^2\Delta}
{R(R-\Delta )}+V(R),
\end{equation}
where $V(R)$ decreases as $\exp (-2R/a)$ as $R\gg a$.
It is important, that $\Delta$ depends on $R$ and $\Delta\to 0$ as
$(R/a)\to\infty$.
In other words, the dipole moment caused by the atom polarization
decreases as $(R/a)\to\infty$. It is very difficult to specify the dipole
moment dependence on $R$. Nevertheless, we can say that $R^{-6}$--term in
the L--J--potential is not correct in liquid media. As to the pair of
$^4$He atoms in vacuum, $R^{-6}$ term is not correct due to the dependence
$\Delta (R)$ (that is unknown unfortunately), and all of these properties
are inherent for $^4$He as the most compact inert gas.

\vspace{1cm}

D. B. Baranov and V. S. Yarunin are thankful to Russian Foundation
for Basic Research project 00-02-16672 for support.

\end{document}